\documentclass[aps,twocolumn,showpacs,preprintnumbers,amsmath,amssymb]{revtex4}

\usepackage{graphicx}

\begin{document}

\title{ Primary Photoexcitations and  the Origin of the Photocurrent \\ in Rubrene Single Crystals}

\author{Hikmat Najafov}
\author{Ivan Biaggio}
\affiliation{Department of Physics and Center for Optical Technologies, 
Lehigh University, Bethlehem, PA 18015}

\author{Vitaly Podzorov}
\author{Matt Calhoun}
\author{Michael E. Gershenson}
\affiliation{Department of Physics and Astronomy, Rutgers University, Piscataway, New Jersey 08854}

\date{\today}

\begin{abstract}
By simultaneously measuring the excitation spectra of  transient 
luminescence and  transient photoconductivity after picosecond 
pulsed excitation in rubrene single crystals we show that free 
excitons are photoexcited starting at photon energies above 2.0 eV. We observe a competition between photoexcitation 
of free excitons and photoexcitation into vibronic  states 
that subsequently decays into free carriers, while  
molecular excitons are instead formed predominantly through the 
free exciton. At photon energies below 2.25 eV, free charge 
carriers are only created through a long-lived intermediate state 
with a lifetime of up to 0.1 ms and no free carriers appear during 
the exciton lifetime.
\end{abstract}

\maketitle 

In organic molecular crystals (OMCs) the small interaction between neighboring molecules naturally
leads to the expectation that photoexcitation should  result in  localized transitions within individual molecules rather than in the delocalized interband transitions typical of covalently bound semiconductors.  In fact, the photoinduced generation of free carriers in OMCs was thought for a long time to predominantly occur through the intermediary
of localized excitonic states that would autoionize to create free carriers \cite{Pope84Pope99,Silinsh96}.

However, by probing the existence of free carriers through their effect on
THz optical pulses it has recently become possible to detect the signature of a free carrier
density that appears to arise immediately after photoexcitation. As an example it has been possible to observe the appearance of free carriers within 0.5 picoseconds after photoexcitation in pentacene \cite{Hegmann02Ostroverkhova05}. These experiments seem to imply that free carriers can indeed be directly photoexcited in OMCs.

The question as to what is the most appropriate description of charge-carrier photogeneration
in OMCs  has not been answered yet in general, and there is an ongoing debate in the literature
which also includes the nature of photoexcitations in conjugated polymers.
In addition, the variety of excitonic states that can be excited in OMCs can make
the photoexcitation picture in these materials quite complex \cite{Pope84Pope99,Silinsh96,Matsui90,Davydov71,Cordella03}. The most obvious excitonic state
is the equivalent of a Frenkel exciton, where photoexcitation
leads to the promotion of the ground state electron onto one of the vibronic sublevels of the
first excited state of a molecule. But it has also been shown that a purely electronic
``frozen lattice'' transition can result in what has been called a free exciton, and that local lattice deformations can lead to self-trapped excitons \cite{Matsui90,Toyozawa83,Silinsh96}.

In this context, it is important to note that a spectral investigation can provide more relevant information on the nature of the primary photoexcitation than only time-resolved studies.
Besides the obvious fact that   different photon energies can create different excitations, the energetic structure of any primary excitation that is the main contributor to a given physical observable will be necessarily reflected in the excitation spectrum of that observable. This in particular also applies to the photocurrent amplitude. Its excitation spectrum would be different depending if the principal mode of initial excitation is, e.g., a localized exciton or a band-to-band transition. Moreover, this applies even in the case where the relaxation from the initial excitation to other intermediary states is so fast that it cannot be directly detected with time-resolved techniques.

In this Letter, we investigate the primary photoexcitation mechanisms  in Rubrene single crystals. We observe a delayed photocurrent after pulsed illumination that directly indicates the existence of a long-lived intermediate state responsible for the creation of free charge carriers, and  by an analysis of photoluminescence and photocurrent excitation spectra we identify two different kinds of primary photoexcitation mechanisms, including a purely electronic transition that can be described as a free exciton. 
Rubrene has shown one of the highest carrier mobilities 
observed to date in organic crystals at room temperature \cite{Podzorov04}, 
can be grown with a high purity, and has been successfully used 
as the transport layer in organic field-effect transistors (FETs) 
\cite{Boer04}. While its high mobility is well characterized in the FET 
geometry \cite{Podzorov03}, and theoretical analysis of its transport properties
have appeared \cite{daSilva05}, to our knowledge no experimental characterization of its
optical and photoexcitation properties has been published yet.

In our experiments, a 1 mm thick rubrene sample was illuminated with 20 ps long laser 
pulses at a repetition rate of 10 Hz. A   constant voltage of 700 V was applied to two surface silver electrodes 
 spaced 4 mm apart along the \textit{b}-axis of the crystal. The wavelength 
of the light pulses was tuned continuously in the spectral 
range between 420 and 700 nm. The illuminated  area of 1.5 
mm$^{2}$ was in the middle between the two electrodes. The pulse 
fluence was kept at 2 $\mu$J/cm$^{2}$ at all wavelengths. This fluence is in the range where the  photocurrent  response grows linearly with the pulse energy. The dynamics of 
the photoresponse was monitored by a 2 GHz/s sampling oscilloscope  
and averaged 150 times. Because of the small 
photocurrent necessitating a large shunt resistor, the response 
time of the detection was limited to $\sim 5 \ \mu$s. 
The experimental set-up allowed for the simultaneous detection 
of both the photoinduced luminescence transient and the photocurrent 
transient. It is important to note that the absorption in rubrene is already larger than
about 50 cm$^{-1}$ around 600 nm, so that practically all the energy in our laser pulses is
absorbed for wavelengths of 600 nm and below. The constant amount of deposited energy
guarantees that all our excitation spectra directly correspond to the efficiency with which absorbed
energy is converted into luminescence or current.

\begin{figure}[t]
         \includegraphics[width=8.5cm]{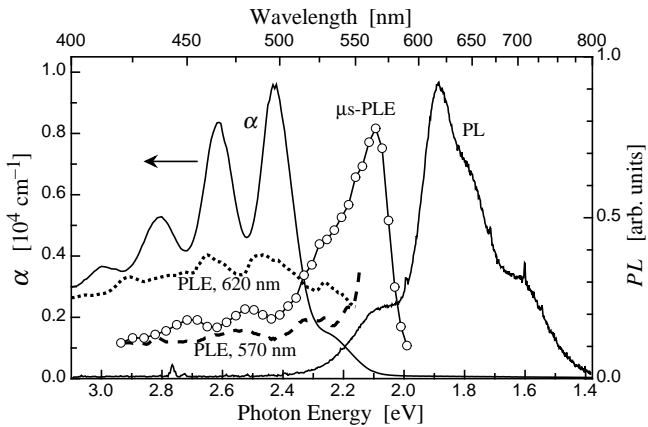}
\caption{ $b$-polarized absorption spectrum ($\alpha$), photoluminescence spectrum (PL),
cw luminescence excitation spectrum (PLE) monitored at 570 nm (dashed line) and 620 nm  (dotted line),
and excitation spectrum of the microsecond luminescence transient induced by $b$-polarized light (white circles).}
\label{f1}
\end{figure}

The photoluminescence (PL) emission spectrum of rubrene  determined 
under continuous wave (cw) excitation is shown in Fig.~\ref{f1}, together with  the rubrene absorption spectrum, which shows clear vibronic bands with a separation of about 0.18 eV. This agrees with the largest Raman line at 1430 cm$^{-1}$ that we observed in Rubrene crystals, and that can be  attributed to a C-C stretching vibration (see, e.g., Ref.\ \onlinecite{Cordella03}).
One notes a typical mirror-like relationship between the vibronic bands of the PL emission at 615 nm and above and those of its cw excitation spectrum and of the absorption, with the weaker, higher energy PL band near 570 nm plays the role of a zero-phonon line. The excitation spectrum of the 570 nm  band, however, is markedly different from the excitation spectrum of the PL emitted
at all wavelengths above 600 nm, while it would need to have the same excitation spectrum if it
was a zero-phonon transition belonging to the same family of vibronic levels. The excitation spectrum of the 570 nm band is weaker than that of the 615 nm band for all shorter wavelengths
down to the UV spectral region that is not shown in Fig.\ \ref{f1}. 
It follows that the 570 nm emission band   must be due to a
different emitting species. Among other reasons that we will discuss below, the overlap of its excitation spectrum with its emission spectrum
leads us to associate it with a purely electronic, ``frozen lattice'' transition, which
can be assigned to a free exciton \cite{Matsui90,He04Cazayous04}. 
The 615 nm band and its vibronic structure, on the other hand, can be assigned to a
(probably self-trapped) molecular exciton \cite{Matsui90}.

\begin{figure}[t]
         \includegraphics[width=8.5cm]{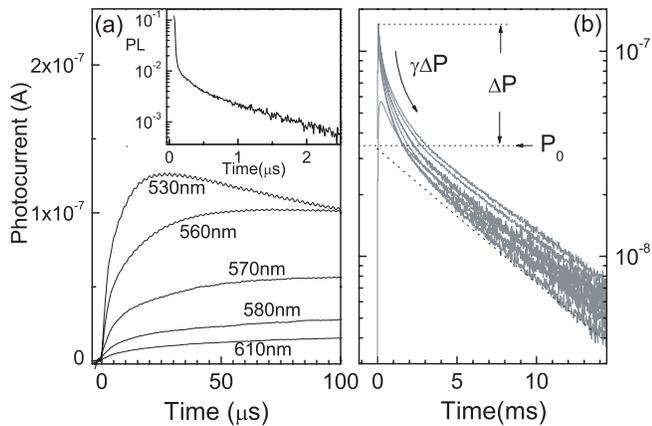}
\caption{
(a) Detail of the photocurrent build-ups for long wavelength excitation. The 
Inset shows the time dynamics of the photoluminescence 
after the pulsed excitation.
(b) Photocurrent transients in the shorter wavelength region, taken every 20 nm between the  excitation wavelengths of 570 nm (lowest curve) and 430 nm.  The parts of the decays that correspond to the amplitudes
$\Delta P$ and $P_0$ are indicated. 
}
\label{f2}
\end{figure}

A striking result, which contrasts with the cw results discussed above, is that  we obtain essentially  the same excitation spectrum at all monitoring wavelengths when observing the PL transient after
pulsed excitation.  The PL dynamics is characterized by an exponential decay with a  time constant of $1.05 \pm 0.2 \ \mu$s that follows a 
fast transient in the first 10 nanoseconds after illumination (see inset in Fig.\ \ref{f2}b).  Fig. \ref{f1} shows the
excitation spectrum obtained by monitoring the PL at 620 nm and 1 $\mu$s after photoexcitation.
This spectrum is very similar to  the cw excitation spectrum of the 570 nm band in the wavelength range where they can be compared. Both spectra show evident \textit{minima} at the photon energies corresponding to the first vibronic resonances in the absorption near 2.2, 2.4, and 2.6 eV. 
These features are  consistent with our assignment of the 570 emission band 
to a free exciton. The excitation to a free exciton continuum, according 
to Toyozawa's description \cite{Toyozawa83}, is expected to be a relatively 
smooth function of the photon energy when compared to the excitation to vibronic  
levels. Since  the total number of absorbed photons is always the same the two 
effects compete: less free excitons are created when more photons 
are lost to excitation of vibronic levels. 
But since the transient PL arising from the molecular exciton has the same excitation spectrum as
the free exciton, it follows that the molecular excitons that decay radiatively on the microsecond time scale are predominantly created by a transition from a free-exciton state.
Whatever excitation is left by the illumination 
pulse in the molecular vibronic levels, it \textit{does not} 
directly decay by multiphonon relaxation to the ground state 
of the molecular exciton. From the fact that photoexcitation in
correspondence with the vibronic levels \emph{does} lead to an increased luminescence when
measured under cw conditions, it follows that there must be an important, slow contribution to
the creation of molecular excitons that is enhanced by absorption into vibronic levels and that dominates the PL  in cw experiments while emitting a too low photon flux to be 
detected as a transient. We argue that such a photoluminescence 
must be due to photoexcited electrons and holes that meet at 
a later time to form a molecular exciton and radiatively recombine. 
The main conclusion is that when vibronic levels are photoexcited, 
the system has a higher probability of generating free carriers 
than of reaching  the ground state of a  
molecular exciton.

\begin{figure}[h]
         \includegraphics[width=8.5cm]{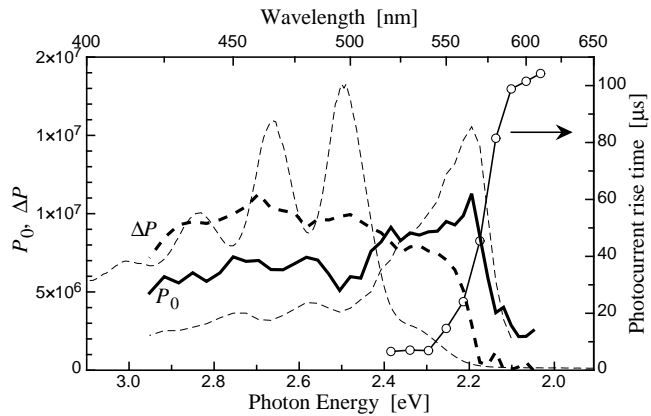}
\caption{
Spectral dependencies of the amplitudes of the quadratic 
($\Delta P$) and  and exponential 
($P_{0}$) components of the  photocurrent, together with its exponential rise time 
 (open circles). The absorption 
spectrum and the transient luminescence excitation spectrum of Fig.~\ref{f1} are shown as
a reference (thin dashed lines).}
\label{f3}
\end{figure}

The transient photoconductivity curves that we determined together with the luminescence are shown in  Fig.~\ref{f2} for 
different excitation wavelengths. Since rubrene is known to be 
a hole dominated material we associate the observed photocurrent 
to holes. From this data one can derive a total number of  carriers $P(t)$ that is contributing to the current  from
$P(t) =I(t) d^{2} /e\mu V $, where $I$ is the photocurrent, $d$ is the inter-electrode 
distance, $e$ is the unit charge, $V$ is the applied voltage, 
and  $\mu$ is the hole mobility, which is taken to be 10 cm$^{2}$V$^{-1}$s$^{-1}$ \cite{Podzorov04b}. 
In  Fig.~\ref{f3} we plot separately the excitation spectra corresponding to the amplitude of the initial quadratic decay ($\Delta P$) of the photocurrent and the amplitude of the exponential decay ($P_0$), together with the rise-time of the photocurrent. 
When compared to the number of photons in the laser pulse, the 
peak number of excited holes is about $10^{4}$ times smaller. This 
means that either the quantum efficiency for hole generation 
is of the order of 10$^{-4}$, consistent in order of magnitude with the results of Ref.\ \onlinecite{Podzorov05}.

Apart from an initial transient, which is only significant 
at shorter wavelengths where it is recognized as a quadratic 
(hyperbolic) decay, the photocurrent decays exponentially with 
a millisecond time constant  which shows little variation 
with the wavelength of the pulsed excitation and which must be 
associated to an average effective lifetime of the photoexcited 
carriers. It may be due to the influence of impurities that provide 
recombination through a trap state in the band gap. 
Since this relaxation process does not affect the measurements 
on the shorter time scales and the initial amplitude of the exponential 
decay, we do not discuss it further here.

From the photocurrent spectra in  Fig.~\ref{f3} one sees that the amplitude 
$P_{0} (\lambda )$ of the exponential decay in the free-carrier number increases 
as soon as the photon energy approaches the free exciton band 
at 570 nm and peaks close to the maximum of the free exciton transition. $P_{0} (\lambda )$ dominates the number of photoinduced carriers until the amplitude $\Delta P(\lambda )$ of the quadratic decay becomes  significant. This happens when the linear 
absorption is so strong that it confines the excitation close 
to the surface of the sample, creating a larger carrier density 
and hence a stronger quadratic recombination. Since the total 
number of absorbed photons is constant, whenever 
$\Delta P(\lambda )$ increases (especially in correspondence with the vibronic absorption 
lines), $P_{0} (\lambda )$ decreases. By fitting the photocurrent time-dynamics at the 
shorter wavelengths and calculating the carrier density using 
the absorption data plotted in  Fig.~\ref{f1} and the total number of 
carriers in  Fig.~\ref{f3}, it is possible to determine a quadratic recombination 
coefficient of  $4.0(\pm 1.5)\times 10^{-10} $
 cm$^{3}$s$^{-1}$. This value is remarkably constant in the wavelength 
interval between 510 and 420 nm where the quadratic recombination 
is clearly observed.  The increased quadratic recombination  leads to an effective decrease of the number of carriers left behind in the crystal after the pulsed illumination for optical
energies corresponding to the vibronic resonances. But under cw excitation  quadratic recombination does not play
a role, and excitation in coincidence with the vibronic resonances will lead to an enhanced creation of free electron and hole pairs that will then be able to recombine at a later time to form molecular excitons, and consequently lead to the increased PL
efficiency observed under cw excitation.

A stunning feature of the photocurrent dynamics appears for
excitation energies below 2.25 eV: the illumination pulse
does not immediately create any free carriers and the photocurrent grows quite slowly after photoexcitation, long after the microsecond decay of the photoluminescence. This
means that free carriers start being  generated  after the 
excitons responsible for the PL have recombined. The build-up 
time of the photocurrent has an almost constant value of the 
order of 0.1 ms in the wavelength range between 610 and 590 nm 
(2.04-2.10 eV). This time must correspond to the lifetime of 
an intermediate state that autoionizes into electrons and holes 
and that must be formed starting from the free exciton. 
At higher excitation
energies, the rise time of the current, and hence the intermediate
state lifetime, shortens until, at excitation 
energies larger than 2.25 eV, we cannot resolve the build-up time 
of the photocurrent anymore. For these higher excitation energies  we cannot exclude an instantaneous  contribution to the photocurrent through  creation of delocalized 
electrons and holes, especially when the excitation energy 
corresponds to the vibronic resonances, as mentioned above.
But in the range of photoexcitation energies between 2.0 and 2.25 
eV, the general coincidence  between the excitation spectrum $P_{0} (\lambda )$ and the free exciton spectrum indicates that the free exciton must play a role in the creation of electrons and holes.
In this spectral region we have very clear proof that photoexcitation does not create 
any electron-hole pairs either directly or through ionization 
of the free exciton. Instead, the free-exciton ultimately decays 
into a longer-lived intermediate state that then  
autoionizes into free carriers. Such an intermediate state may be 
associated with the triplet exciton that can be formed through 
intersystem crossing in many organic systems \cite{Kleinerman62,Jundt95,Agostini83}.

In conclusion, we have shown that the primary photoexcitation in rubrene 
single crystals consists in free excitons starting at photon energies 
above $\sim 2.0$ eV, with also some excitation of vibronic levels 
at the higher photon energies of 2.5, 2.67, and 2.85 eV. The 
free exciton can radiatively recombine to emit a PL band centered 
around 570 nm, can convert into a  molecular exciton 
that radiates the PL band that peaks at 620 nm, or can lead (possibly 
through the molecular exciton) to an intermediate 
state with a lifetime of up to 0.1 ms which then self-ionizes into electrons and holes.
Important highlights of this work are  (1) the observations that
the excitation 
in the vibronic resonances does not decay to the ground state 
of a  molecular exciton but leads instead to delocalized 
carriers that only later meet to form a molecular exciton and 
radiatively recombine; and (2) the clear and direct evidence, provided by the observed delayed onset of the photocurrent, that it is impossible to obtain free carriers directly  at optical excitation energies lower than 2.25 eV, where free carriers are exclusively 
formed by autoionization of a long-lived intermediate state 
originating from the initial creation of a free exciton. We are not aware of any other observation  of such a delayed photocurrent onset as we have seen in Rubrene in this work.
Finally, the observation of a free exciton and  the fact that it can be photoexcited over a wide spectral range raise the question as to weather the THz absorption signal observed in Refs.\ \onlinecite{Hegmann02Ostroverkhova05} in pentacene may not be due to transitions induced by the THz field between the closely spaced sublevels of the free exciton \cite{Toyozawa83,He04Cazayous04}, similar to the microwave transitions that have been observed to occur between free exciton sublevels in other organic materials \cite{Agostini83}, which brings us back to the initial question about the nature of the primary photoexcitation in OMCs. Additional insights into this questions and into the electronic states of Rubrene single crystals will be obtained by a further investigation of the
effects presented here, such as of their temperature quenching properties.

Acknowledgments: this work at Rutgers has been supported by the NSF grants DMR-0405208 and
ECS-0437932.

\end{document}